\documentclass[prl,twocolumn,superscriptaddress,a4paper,showpacs]{revtex4}
\usepackage{amsmath,graphicx}
\usepackage{color}
\begin{document}

\title{Anisotropic spin relaxation in graphene}
\author{N. Tombros}
\affiliation{Physics of Nanodevices,}
\author{S. Tanabe}
\affiliation{Physics of Nanodevices,}
\author{A. Veligura}
\affiliation{Physics of Nanodevices,}
\author{C. Jozsa}
\affiliation{Physics of Nanodevices,}
\author{M. Popinciuc}
\affiliation{Molecular Electronics,
Zernike Institute for Advanced
Materials, Rijksuniversiteit Groningen, Nijenborgh 4, 9747 AG
Groningen, The Netherlands}
\author{H. T. Jonkman}
\affiliation{Molecular Electronics,
Zernike Institute for Advanced
Materials, Rijksuniversiteit Groningen, Nijenborgh 4, 9747 AG
Groningen, The Netherlands}
\author{B. J. van Wees}
\affiliation{Physics of Nanodevices,}
\date{\today}
\begin{abstract}
\textbf{Spin relaxation in graphene is investigated in electrical
graphene spin valve devices in the non-local geometry. Ferromagnetic
electrodes with in-plane magnetizations inject spins parallel to the
graphene layer. They are subject to Hanle spin precession under a
magnetic field $B$ applied perpendicular to the graphene layer.
Fields above 1.5 T force the magnetization direction of the
ferromagnetic contacts to align to the field, allowing injection of
spins perpendicular to the graphene plane. A comparison of the spin
signals at $B$ = 0 and $B$ = 2 T shows a 20 $\%$ decrease in spin
relaxation time for spins perpendicular to the graphene layer
compared to spins parallel to the layer. We analyze the results in
terms of the different strengths of the spin orbit effective fields
in the in-plane and out-of-plane directions.}
\end{abstract}
 \maketitle

%%%%%%%%%%%%%%%%%%%%%%%%%%%
%    Experiment           %
%%%%%%%%%%%%%%%%%%%%%%%%%%%
The discovery of the anomalous quantum Hall effect in graphene
\cite{Novoselov, Zhang} triggered an avalanche of theoretical and
experimental work on this new system. Spintronics is one of the
fields which has great expectations for this material. Spin qubits
\cite{Loss} and many other spintronic devices based on graphene
could become available due to the fact that in intrinsic graphene
spins are expected to relax very slowly
\cite{Kane,Huertas,YaoSpinOrbit,MinSpinOrbit}. The reason behind
this is the low hyperfine interaction of the spins with the carbon
nuclei (only 1 $\%$ of the nuclei are C13 and have spin) and the
weak spin-orbit (SO) interaction due to the low atomic number.

Recent experiments show spin transport in graphene up to room
temperature
\cite{TombroGraphene,HillGraphene,ChoGraphene,NishiokaGraphene,OhishiGraphene},
with spin relaxation lengths of 2 $\mu$m and relaxation times around
150 ps \cite{TombroGraphene}. Such relatively short relaxation times
suggest an important role of SO interaction. There are two relevant
mechanisms for SO interaction in graphene \cite{Fabian}. In the
Elliott-Yafet (EY) mechanism, spin scattering is induced by electron
(momentum) scattering from impurities, boundaries and phonons. The
D'yakonov-Perel' (DP) mechanism results from SO terms in the
Hamiltonian of the clean material. Here electrons feel an effective
magnetic field, which changes in direction every time the electron
scatters to a different momentum state, resulting in random spin
precession. In principle the two mechanisms can be distinguished by
their different dependence on the momentum scattering time $\tau$
\cite{Fabian}. In our experiments in graphene we are not able to
change $\tau$ considerably, making the distinction between
Elliott-Yafet and D'yakonov-Perel' mechanisms difficult. However we
can obtain valuable information about the SO interaction by
investigating the anisotropy of spin relaxation. First we note that
the transverse ($T_2$) and longitudinal ($T_1$) spin relaxation
times are expected to be the same for the parameters of our system
\cite{Fabian}. Therefore, as in metals, a single spin relaxation
time $T$ = $T_1$ = $T_2$ can be used. However, due to the
2-dimensionality, $T$ can have a different value for injected spins
parallel ($T_{\parallel}$) or perpendicular ($T_{\perp}$) to the
graphene plane. For example, if the SO interaction is of the Rashba
or Dresselhaus type then the SO effective fields are exclusively in
the graphene plane and calculations show that this should result in
anisotropic spin relaxation in which $T_{\perp}$ = 1/2 $
T_{\parallel}$ \cite{Fabian}. On the other hand, if the SO effective
fields pointing out of the graphene plane dominate, we expect
$T_{\perp} >> T_{\parallel}$. Here we will directly compare the spin
relaxation times in the parallel and perpendicular directions,
measured under identical experimental conditions.

\begin{figure}[htb]
\begin{center}
\includegraphics[width=7cm]{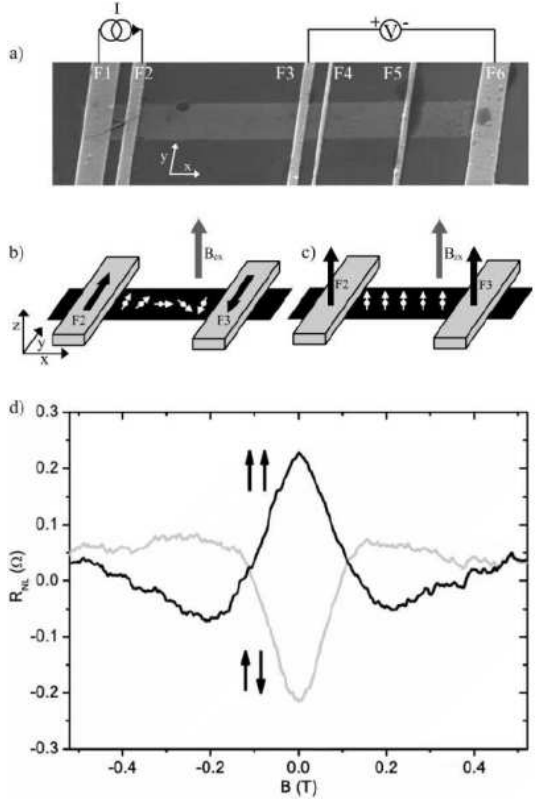}
\end{center}
\caption{Spin transport in graphene a) A SEM picture of a single
layer of graphene contacted by 6 cobalt electrodes (sample $A$).
Spins traveling a distance of 5 $\mu$m, from cobalt electrode $F$2
to $F$3, are probed using the 'non-local' geometry. The voltage
circuit ($F$3-graphene-$F$6) is completely separated from the
current circuit ($F$1-graphene-$F$2), b) Hanle type spin precession
experiment, the magnetization of the spin injector $F$2 is set
antiparallel to the magnetization of spin detector $F$3. Spins are
injected parallel to the graphene plane.
 c) Application of a strong external magnetic field ($\sim$1.4T) perpendicular to the graphene layer
 results to injector and detector magnetizations aligned parallel to
 the external magnetic field. Spins are injected perpendicular to the graphene plane.
 d) Spin precession in case of parallel ($\uparrow\uparrow$, black curve) and antiparallel
 ($\uparrow\downarrow$, grey curve) magnetizations. } \label{fig:device_local_nonlocal}
\end{figure}

Our experiments are performed using the four terminal "non-local"
technique (Fig. 1a). Here the charge current path can be fully
separated from the voltage detection circuit. The non-local
technique is less sensitive to device resistance fluctuations and
magnetoresistances (such as Hall effects), as compared to the
standard two-terminal spin valve technique. This allows the
detection of small spin signals, in our case as small as a few
m$\Omega$ (Fig. 2a). Fabrication of the devices is done as in Ref.
8. Using the "Scotch tape" technique \cite{Geim} graphene layers
were deposited on an oxidized (500 nm) heavily doped Si wafer.
Calibrations by Raman spectroscopy in combination with optical
microscopy and atomic force microscopy show that our samples are
single graphene layers. We evaporate a thin layer of aluminum
(6$\AA$) on top of the graphene layer at 77 K and let it oxidize
using pure O$_2$, to form an Al$_2$O$_3$ barrier. These barriers
very likely contain pinholes \cite{TombroGraphene}, nevertheless
spin injection efficiencies of 10 $\%$ have been observed.
Conventional electron beam lithography and e-beam evaporation of 50
nm of Co (at $10^{-6}$ mbar) are used to define the ferromagnetic
cobalt electrodes. The electrodes have different widths to assure
different switching fields \cite{Jedema2}. The experiments are
performed at a temperature of 4.2K and we use magnetic fields up to
4.5 Tesla. A standard a.c. lock-in technique is used with currents
in the range 1-20 $\mu$A.

Spin precession measurements are performed on two samples (graphene
width $W$ = 1.2 $\mu$m) for electrode spacings $L$ = 5 $\mu$m
(sample $A$), 0.5, 2 and 4$\mu$m (sample $B$). To perform the Hanle
spin precession experiments we first apply a magnetic field in the
$y$-direction to prepare the magnetizations of the electrodes in a
parallel or antiparallel orientation (Fig. 1a). Then this field is
removed and a $B$-field in the z-direction is scanned (Fig. 1b)
\cite{Jedema2}. An example of the resulting spin precession is
depicted in Fig. 1d (sample $B$), for the parallel and antiparallel
magnetizations of the spin injector and spin detector cobalt
electrodes. The spins are injected parallel to the graphene plane
and are precessing while diffusing towards the spin detector
situated at a distance $L$ = 4 $\mu$m from the injector. At B$_z$
$\sim$ 0.2 T the average precession angle is 180 degrees, resulting
in a sign reversal of the spin signal. The magnitude of the signal
at B$_z$ = 0T (0.2 $\Omega$) is small compared to the signals
measured in our previous work, which was in the order of 2 $\Omega$
for these spacings \cite{TombroGraphene}. This is directly related
to the low contact resistances $R_c$ (1-2 k$\Omega$) found between
the cobalt electrodes and the graphene/Al$_2$O$_3$, which are a
factor 5 to 10 smaller than in Ref. \cite{TombroGraphene}. In this
study, the contact resistance $R_c$ is equal or smaller to the
typical square resistance of the graphene layer $R_{sq}$ and this
results in the reduction of the injection/detection efficiencies and
also provides an extra path for spin relaxation at the ferromagnetic
contacts \cite{conductivitymismach}. This is taken into account
later in the fitting of the spin precession measurements with
solutions of the 1-dimensional Bloch equations \cite{Jedema2} which
describe the combined effect of diffusion, precession and spin
relaxation in the system. From the fit, the intrinsic spin
relaxation time in graphene can be extracted. The relative
importance of the contacts is given by the parameter $R$ = $W\cdot R
_c /R _{sq}$, where $W$ is the width of the graphene layer. If the
spin relaxation length is in the $\mu$m range then the model shows
that for $R$ $>>$ 10$^{-5}$ m the contacts do not induce extra spin
relaxation. On the other hand for $R$ $<<$ 10$^{-5}$m the amplitude
($A$) of the spin signal has quadratic dependence in $R$ ($A \sim
R^2$) ( M. Popinciuc $et$ $al$., in preparation).

\begin{figure}[htb]
\begin{center}
\includegraphics[width=7cm]{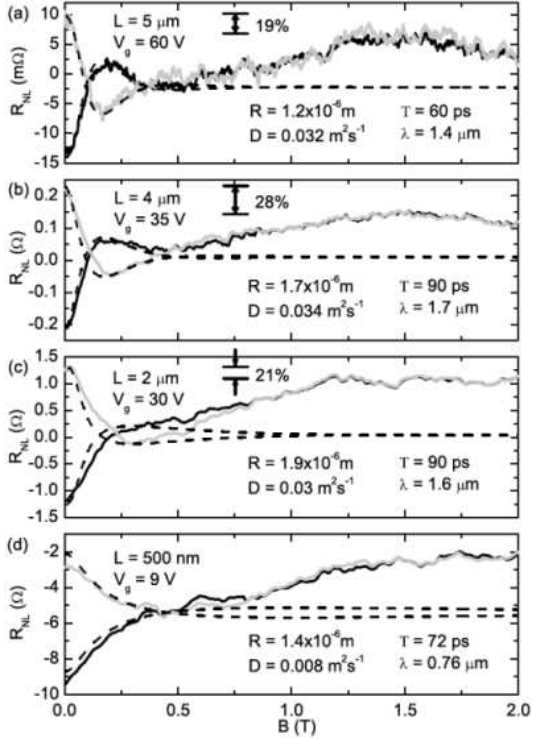}
\end{center}
\caption{Anisotropic spin relaxation at high electron density $n$ a)
$n$ = 2.1 10$^{12}$ cm$^{-2}$. Initially, spins are injected
parallel to the graphene plane having the magnetization of the spin
injector set parallel (gray line) or antiparallel (black line) to
the detector (at a distance $L$ = 5 $\mu$m). From the fits (dashed
line) we extract the diffusion constant $D$ and the relaxation time
$T_{\parallel}$. A magnetic field of $\sim$1.4 T is needed to align
the magnetization of the cobalt electrodes out of their easy
magnetization axis, in this case the spins are injected
perpendicular to the graphene layer having a spin relaxation time
$T_{\perp}$. The spin relaxation $T_{\perp}$ is 19 $\%$ smaller to
$T_{\parallel}$. The small decrease in the nonlocal signal found
between 1.5 and 2 T is attributed to a background due to orbital
magnetoresistance effects. The same experiment has been performed on
sample $B$ for b) $L$ = 4 $\mu$m ($n$ = 3.5 10$^{12}$ cm$^{-2}$), c)
$L$ = 2 $\mu$m ($n$ = 2.8 10$^{12}$ cm$^{-2}$) and d) $L$ = 500 nm
($n$ = 3.5 10$^{12}$ cm$^{-2}$).} \label{fig2}
\end{figure}
%PreceFittingNewModel

In our experiments a background arises in the non-local resistance
which has a quadratic dependency on $B_z$ due to orbital
magnetoresistance effects, in particular around the Dirac point. At
high electron densities this background is usually small. We
therefore start by applying a gate voltage on samples $A$ and $B$ to
allow us to investigate the spin dynamics at a high electron density
n$\sim3.0$ $10^{16}m^{-2}$(Fig. 2). From the fitting procedure both
the diffusion constant $D$ and the spin relaxation time
$T_{\parallel}$ can be obtained. For $L$ = 2, 4 (Sample B) and
5$\mu$m (Sample A) we obtain a diffusion constant of 3$\cdot$
10$^{-2}m^2s^{-1}$ and spin relaxation times of $T_{\parallel}$ = 60
ps for sample $A$ up to 90 ps for sample $B$, corresponding to spin
relaxation lengths $\lambda$=$\sqrt{DT_{\parallel}}$ of 1.4 up to
1.8 $\mu$m, respectively, comparable to the values found in Ref.
 \cite{TombroGraphene}. Increasing the magnitude of $B_z$ results in
a rotation of the magnetization of the cobalt electrodes out of the
plane, towards the magnetic field direction. This can already be
seen in Fig. 1d where the rotation of the magnetization induces an
asymmetry of the spin signal at 0.5 T. A magnetic field of 1.4 to
1.8 T is needed to fully align the magnetization of the cobalt
electrodes in the $z$-direction (Fig. 1c)
\cite{Jedema2,BrandsaSaturationField}. This is a special situation
as injected spins are now perpendicular to the graphene layer and
will relax with a time $T_{\perp}$ which is not necessarily  the
same as $T_{\parallel}$. If the anisotropy in the spin relaxation is
large then the amplitude of the non-local spin signal at B = 0 T
should be very different from the signal at $\sim$1.8 T. The high
sensitivity of this method comes from the fact that the amplitude of
the spin signal depends exponentially on $\lambda$. For example, for
$L$ = 5 $\mu$m and $\lambda$ = 1.4 $\mu$m a decrease by 10 $\%$ in
$\lambda$ due to anisotropic spin relaxation should result in a 40
$\%$ decrease in spin signal amplitude. In this example a 10 $\%$
decrease in $\lambda$ corresponds to a 20 $\%$ ($100\%
\cdot(T_{\parallel}-T_{\perp})/T_{\perp}$) decrease in $T$. In Fig.
2 the decrease in the magnitude of the spin signal for $L$ = 2, 4
and 5 $\mu$m corresponds to a spin relaxation time $T_{\perp}$ being
20 $\%$ smaller than $T_{\parallel}$. Clearly, our devices show
anisotropic spin relaxation in graphene at high electron densities.
Of interest is to investigate if the same conclusion holds for spins
injected in graphene at the charge neutrality point. For $L$ = 2, 4
and 5 $\mu$m, close to the Dirac point, orbital magnetoresistance
effects induce a large background, increasing quadratically in
$B_z$. This background is not only monotonic increasing, it also
contains non-periodic fluctuations as a function of $B_z$ with an
amplitude equal or larger than the spin signal. This effect, in
combination with the large suppression of the spin signal amplitude
at the Dirac point, for $L$ = 2, 4 and 5 $\mu$m, does not allow us
to investigate in precision the spin anisotropy. However, we were
able to perform the experiment for the $L$ = 0.5 $\mu$m spacing in
which the spin signal is relatively large. (Fig. 3c) Here,
application of a gate voltage of -76.5 V (Fig. 3a) allows us to
investigate the spin dynamics at the Dirac point. Clearly, the
non-local resistance at 2T is smaller than the resistance at $B_z$ =
0T showing similar anisotropic spin relaxation behavior as for high
electron densities.

\begin{figure}[htb]
\begin{center}
\includegraphics[width=7cm]{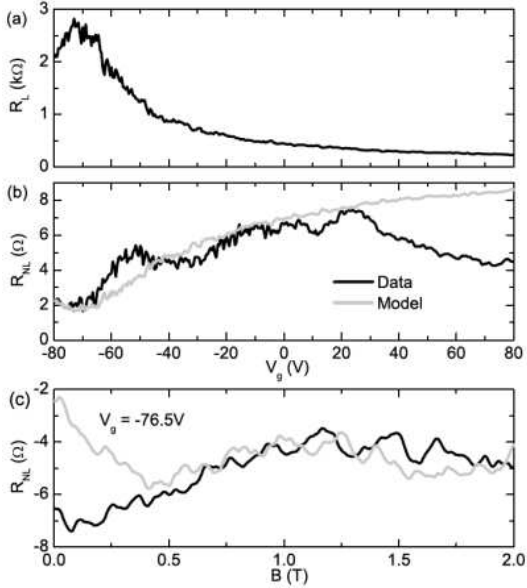}
\end{center}
\caption{Anisotropic spin relaxation at the Dirac point for $L$ =
500 nm a) Gate voltage dependence of the graphene resistance between
electrodes $F_2$ and $F_3$ (Fig. 1b). The charge neutrality point is
found at V$_g$ $\approx$ -75 V. b) The nonlocal resistance for spins
injected parallel to the graphene layer (black line). This is
defined as the difference in the signal obtained for injector and
detector magnetizations set parallel (R$_{NL,\uparrow\uparrow}$) and
the signal for injector and detector set to antiparallel
(R$_{NL,\uparrow\downarrow}$). Our model (gray line) which takes
into account the finite contact resistance gives a qualitatively
good fit to the data for V$_g$$<$30V. c) Application of a magnetic
field $B$ perpendicular to the graphene plane gives similar results
as found in Fig. 2. A spin relaxation for spins injected
perpendicular to the graphene layer is found to be smaller than the
spin relaxation of the spins injected parallel to the layer.}
\label{fig3}
\end{figure}
%A_Hall1_graphs23102007

We can now estimate the effective magnetic field which the electrons
feel, assuming the D'yakonov-Perel' mechanism. The electron scatters
to a different momentum state after a time $\tau$ which results in a
precession angle of the spin $\Delta\omega = \omega_p\tau$. Here,
$\omega_p$ is the precession frequency of the spin. The number of
scattering events necessary to induce an angle of 2$\pi$ is
$\sqrt{T/ \tau}$. Using $T$ = 100 ps and $\tau\sim$ 30 fs we obtain
$\omega_p\sim10^{12}s^{-1}$. Therefore, in the D'yakonov-Perel'
mechanism, the precession frequency corresponds to a magnetic field
of about 5 Tesla.

We now check if the transverse ($T_2$) and longitudinal ($T_1$)
relaxation times for spins in the graphene plane are the same. For
this, $\lambda$ (=$\sqrt{DT_1}$) is extracted from the spin signal
dependence $L$ (Fig. 4). Care has to be given to the strong
suppression of the spin signal amplitude as we approach the Dirac
point, found at 4.2 K and as well at room temperature. Our model
takes into account the spin relaxation at the contacts and fits
qualitatively well the data (Fig. 3b). Earlier work did not show
this strong effect due to the fact that in those samples the contact
resistances were large enough to not to influence the spin dynamics.
In Fig. 4 we present the amplitude of the spin signal of sample $B$
as function of the electrode spacing L (= 0.5 , 2 and 4 $\mu$m) for
different values of $R$. In the same figure we present the length
dependence in the signal expected from our model. At high electron
densities $n$ (Fig. 4a) the model gives a spin relaxation of 1.5
$\pm0.2 \mu$m. This value is comparable to the values found from
spin precession measurements (Fig. 2) performed at similar values of
$n$ (and $R$), proving that $T_1\simeq T_2$. The effect on $\lambda$
when we approach the Dirac point is stronger: our model gives
$\lambda$ = 1 $\mu$m for $n\simeq2.0$ 10$^{12}$cm$^{-2}$ (Fig 4 b)
and $\lambda$ = 0.8 $\mu$m for $n\simeq1.0$ 10$^{12}$cm$^{-2}$ (Fig.
4c).

\begin{figure}[t!]
\begin{center}
\includegraphics[width=7cm]{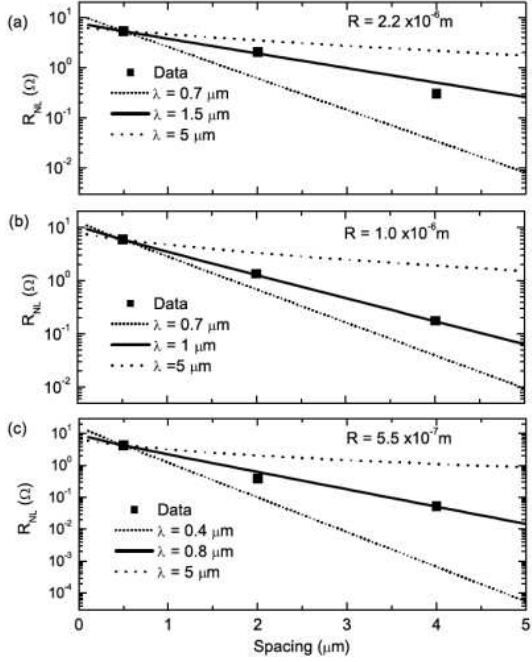}
\end{center}
\caption{Dependence of the spin signal as function of electrode
spacing $L$ for spins injected parallel to the graphene plane. a)
The gate voltage V$_g$ is set in such a way that for $L$ = 0.5 , 2
and 4 $\mu$m we have the same value of $R$ = 2.2 10$^{-6}$
($n\simeq5.0$ 10$^{12}$ cm$^{-2}$, $D$ = 0.04 m$^2$s$^{-1}$). We
obtain a relaxation length of $\lambda$ = 1.5 $\pm$ 0.2 $\mu$m which
is similar to the length extracted from the spin precession
measurements (see Fig. 2). Model calculations for $\lambda=0.7\mu$m
and $\lambda=5\mu$m are shown for comparison.
 b) Moving towards the Dirac point using a gate voltage in such a way
 that we decrease the $R$ value to 1.0 10$^{-6}$ ($n\simeq2$ 10$^{12}$ cm$^{-2}$, $D$ = 0.03 m$^2$s$^{-1}$) has a strong influence in
 $\lambda$, as it decreases to 1 $\mu$m  and c) to
0.8 $\mu$m (for $R$ = 5.5 10$^{-7}$, $n\simeq1.0$ 10$^{12}$
cm$^{-2}$, $D$ = 0.02 m$^2$s$^{-1}$).} \label{fig4}
\end{figure}
%A_Hall1_graphs23102007

Summarizing, we use a non-local measurement geometry to investigate
anisotropic spin relaxation in graphene. Although our experimental
accuracy do not allow us to confirm that T$_1$ and T$_2$ are equal,
they allow us to measure with accuracy the difference between
$T_{\perp}$ and $T_{\parallel}$. At high electron densities, a
decrease up to $\sim$40 $\%$ in the spin signal is found for
injection of spins perpendicular to the graphene layer compared to
injected spins parallel to the graphene. This corresponds to a spin
relaxation $T_{\perp}$ almost 20 $\%$ smaller to $T_{\parallel}$.
This spin anisotropy is expected for a 2-D system were spin-orbit
fields in plane dominate the spin relaxation. As a next step we
suggest to investigate the dependence of $T_{\perp}$ and
$T_{\parallel}$ on momentum scattering $\tau$ to establish the
relative role of the Elliott-Yafet mechanism compared to the
D'yakonov-Perel' mechanism.

We thank Bernard Wolfs, Siemon Bakker, Thorsten Last for technical
assistance and for useful discussions. This work was financed by
MSC$^{plus}$, NanoNed, NWO (via a 'PIONIER' grant) and FOM (via the
'Graphene-based electronics' program). The work of Shinichi Tanabe
was made possible by an Osaka University Scholarship for Exchange
Study.

\end{document}